\documentclass{article}

\usepackage{arxiv}

\usepackage[detect-all]{siunitx}
\usepackage{amsmath}
\usepackage{amssymb}
\usepackage{nicefrac}
\usepackage{microtype}

\usepackage[utf8]{inputenc}

\usepackage{hyperref}
\usepackage{cleveref}
\usepackage{url}
\usepackage[sorting=none]{biblatex}
\usepackage{graphicx}
\usepackage{doi}

\usepackage[acronym]{glossaries-prefix}
\glsdisablehyper
\newacronym{5g}{5G}{fifth generation}
\newacronym{emf}{EMF}{electromagnetic field}
\newacronym{icnirp}{ICNIRP}{International Commission on Non-Ionizing Radiation Protection}
\newacronym{ices}{ICES}{International Committee on Electromagnetic Safety}
\newacronym{br}{BR}{basic restrictions}
\newacronym{drl}{DRL}{dosimetric reference limits}
\newacronym{rl}{RL}{reference levels}
\newacronym{erl}{ERL}{exposure reference levels}
\newacronym{apd}{APD}{absorbed/epithelial power density}
\newacronym{ipd}{IPD}{incident power density}
\newacronym{ks}{KS}{Kolmogorov-Smirnov}
\newacronym{mae}{MAE}{mean absolute error}
\newacronym{mlp}{MLP}{multilayer perceptron}
\newacronym{moe}{MoE}{mixture of experts}
\newacronym{ci}{CI}{confidence interval}

\title{Standardized Benchmark Dataset for Localized Exposure to a Realistic Source at 10--90 GHz}

\author{
    \href{https://orcid.org/ 0000-0001-5507-5788}{\includegraphics[scale=0.06]{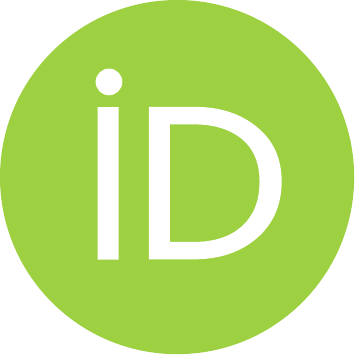}\hspace{1mm}Ante Kapetanovic}\\
	University of Split\\
	Split, Croatia\\
	\texttt{akapet00@gmail.com}\\
	\And
    \href{https://orcid.org/0000-0003-1205-4842}{\includegraphics[scale=0.06]{orcid.pdf}\hspace{1mm}Dragan Poljak}\\
	University of Split\\
	Split, Croatia\\
	\texttt{dpoljak@fesb.hr}\\
    \And
	\href{https://orcid.org/0000-0002-0013-045X}{\includegraphics[scale=0.06]{orcid.pdf}\hspace{1mm}Kun Li}\\
	University of Electro-Communications\\
	Tokyo, Japan\\
	\texttt{li.kun@awcc.uec.ac.jp }\\
}

\renewcommand{\headeright}{A. Kapetanovic et al.}
\renewcommand{\undertitle}{Preprint}
\renewcommand{\shorttitle}{Exposure Assessment Benchmark Dataset}   

\begin{document}
\maketitle

\begin{abstract}
    The lack of freely available standardized datasets represents an aggravating factor during the development and testing the performance of novel computational techniques in exposure assessment and dosimetry research.
    This hinders progress as researchers are required to generate numerical data (field, power and temperature distribution) anew using simulation software for each exposure scenario.
    Other than being time consuming, this approach is highly susceptible to errors that occur during the configuration of the electromagnetic model.
    To address this issue, in this paper, the limited available data on the incident power density and resultant maximum temperature rise on the skin surface considering various steady-state exposure scenarios at \SIrange{10}{90}{\GHz} have been statistically modeled.
    The synthetic data have been sampled from the fitted statistical multivariate distribution with respect to predetermined dosimetric constraints.
    We thus present a comprehensive and open-source dataset compiled of the high-fidelity numerical data considering various exposures to a realistic source.
    Furthermore, different surrogate models for predicting maximum temperature rise on the skin surface were fitted based on the synthetic dataset.
    All surrogate models were tested on the originally available data where satisfactory predictive performance has been demonstrated.
    A simple technique of combining quadratic polynomial and tensor-product spline surrogates, each operating on its own cluster of data, has achieved the lowest mean absolute error of \SI{0.058}{\celsius}.
    Therefore, overall experimental results indicate the validity of the proposed synthetic dataset.
\end{abstract}

\section{Introduction}
    With an advent of personal wireless devices operating in data-intensive regime, expansion to the previously unused bands of radio-frequency spectrum became necessity.
    This is mainly due to the increasing demand for higher data rates and throughput, but also more reliable service connections. The \gls{5g} wireless network -- a novel technology standard for broadband cellular networks, has enabled the realization of these requirements via novel features such as the carrier aggregation, multiple-input and multiple-output technology, and beam-forming~\cite{Andrews2014}.
    
    A significant consequence of the transition to high micro- and millimeter waves has been the growing interest in, as well as concerns about, biological safety due to \gls{emf} exposure~\cite{Wu2015}.
    Consequently, the \gls{icnirp} guidelines~\cite{ICNIRP2020} and IEEE \gls{ices} C.95.1 standard~\cite{IEEE2019} have both recently undergone a major revision.
    The most significant update is the \gls{apd} to be used as the \gls{br}~\cite{ICNIRP2020} or \gls{drl}~\cite{IEEE2019} for localized steady-state exposure above \SI{6}{\GHz}.
    The \gls{rl}~\cite{ICNIRP2020} or \gls{erl}~\cite{IEEE2019} have been given in terms of the \gls{ipd}.
    Both these quantities should be averaged over a square shaped area of \SI{4}{\cm\squared}~\cite{Hashimoto2017}.
    A spatial averaging area of \SI{1}{\cm\squared} for frequencies above \SI{30}{\GHz} should also be used to account for smaller beam diameters~\cite{Foster2016}.

    Two distinct definitions of the spatially averaged \gls{ipd} have been used in the literature and are both established as valid proxies for local maximum temperature rise on the surface of the exposed skin, provided it is out of the reactive near field region of an \gls{emf} source~\cite{Li2021}.
    Namely, the first definition considers normal components of the Poynting vector crossing the surface ($sPD_\text{n}$), whereas the second definition takes into account all components, i.e., the magnitude, of the Poynting vector ($sPD_\text{tot}$).
    
    To develop and test the performance of new computational methods and techniques within the field of \gls{emf} exposure assessment and dosimetry, it is necessary to first collect the high-fidelity numerical data considering different realistic exposure scenarios.
    The collection of such data requires computationally expensive simulations using either in-house code or commercial \gls{emf} software.
    The main issue of the first approach is susceptibility to computational errors, whereas unavailability and limited use is a common pitfall of the latter.
    To this end, we present a standardized comprehensive benchmark dataset generated artificially by using statistical modelling of the existing data on the localized steady-state exposure to a realistic source at \SIrange{10}{90}{\GHz}.

\section{Methods}
\label{sec:methods}

\subsection{Data Collection, Exploration and Processing}
    Data on 4 different \gls{ipd}s ($sPD_\text{n}$ and $sPD_\text{tot}$, averaged on both \SIlist{4;1}{\cm\squared}) and resulting maximum temperature rise have been generated by 6 organizations in the international collaboration study~\cite{Li2021}.
    The main objective was to clarify the causes of numerical calculation errors in dosimetry analyses by inter-comparing computed results between each organizations using its own simulation software and employing various body and antenna models at \SIlist{10;30;60;90}{\GHz}.
    This data were used for the development of the IEEE 2889-2021 standard~\cite{IEEE2021} where it has been shown that either definition of the \gls{ipd}, i.e., $sPD_\text{n}$ or $sPD_\text{tot}$, serves as a valid proxy for maximum temperature rise on the surface provided the tissue is out of the reactive near-field of the antenna.
    
    Here, we use 115 samples of the aforementioned data collected from~\cite{DeSantis2022}, considering exposure to a half-wavelength dipole antenna placed at separation distances of \SIlist{2;5;10;50;150}{\mm} from the body.
    A half-wavelength dipole antenna is resonated with an adjusted length to obtain the maximum possible radiation power and thus realize the worst-case exposure scenario.
    To meet the criteria of synthetic data generation and surrogate modelling that follows, the collected data have been arranged in the tabular form where the first 6 columns are respectively the separation distance, $d$, frequency, $f$ , and peak values of 4 spatially averaged \gls{ipd}s, $sPD_\text{n, 1}$, $sPD_\text{tot, 1}$, $sPD_\text{n, 4}$ and $sPD_\text{tot, 4}$.
    The last column is resulting maximum temperature rise on the surface.
    
\begin{figure}[b]
        \centering
        \includegraphics[width=\textwidth]{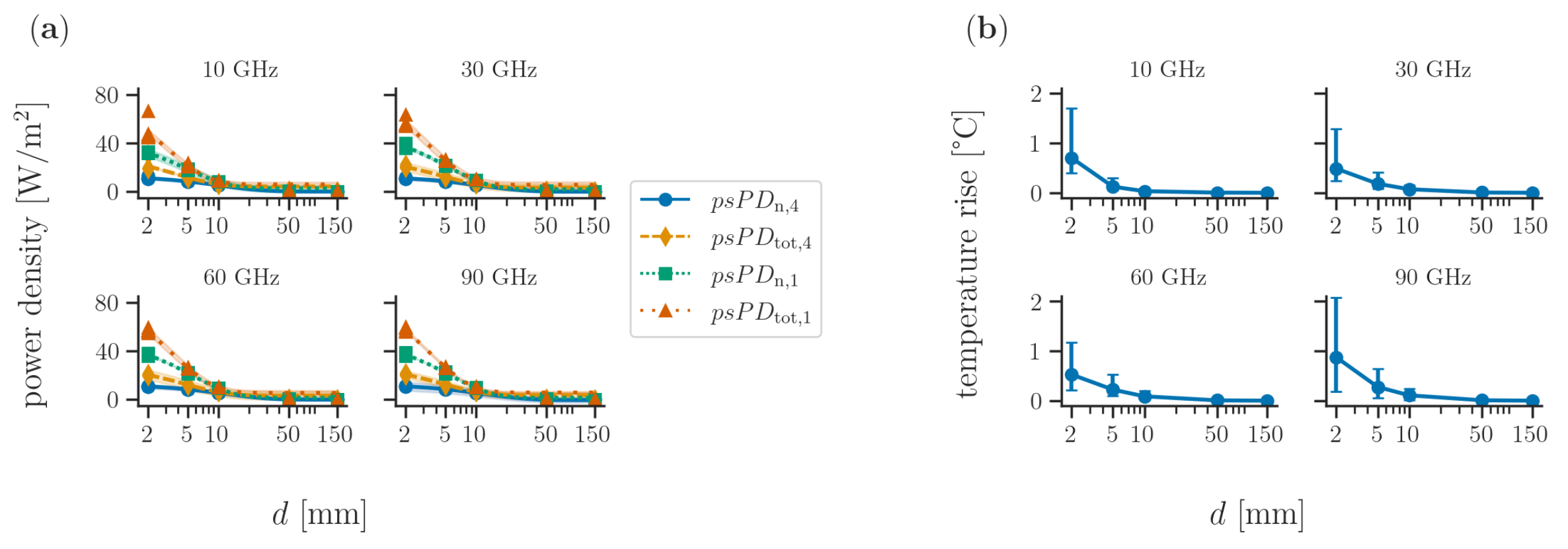}
        \caption{Collected data at \SIlist[list-units=single]{10;30;60;90}{\GHz}: (a) power densities incident on the surface as functions of the antenna-to-tissue separation distance, (b) resulting temperature rise on the surface as a function of the antenna-to-tissue separation distance.}
        \label{fig:1}
    \end{figure}

    In~\cref{fig:1}(a), peak values of spatially averaged \gls{ipd}s as functions of $d$ are shown.
    Both $sPD_\text{n}$ and $sPD_\text{tot}$ decrease monotonically with increasing $d$ where $sPD_\text{tot}$ is greater than $sPD_\text{n}$ across the entire spatial domain.
    All \gls{ipd}s are fitted with the same exponential decay function of varying parameters and placed in the \SI{95}{\percent} confidence interval, showing almost identical behavior, differing only in the power scale which depends exclusively on the \gls{ipd} definition and averaging area.
    Maximum temperature rise in the \SI{95}{\percent} confidence interval as a function of $d$, shown in~\cref{fig:1}(b), follows a similar downward trend.
    It is interesting to notice that the uncertainty is the greatest at $d = \SI{2}{\mm}$, most likely due to discrepancies in numerical methods between organizations, which is especially emphasized in the near-field region.
    Above \SI{5}{\mm}, only marginal differences in estimated temperature rise are present.
    
\begin{figure}[b]
        \centering
        \includegraphics[width=\textwidth]{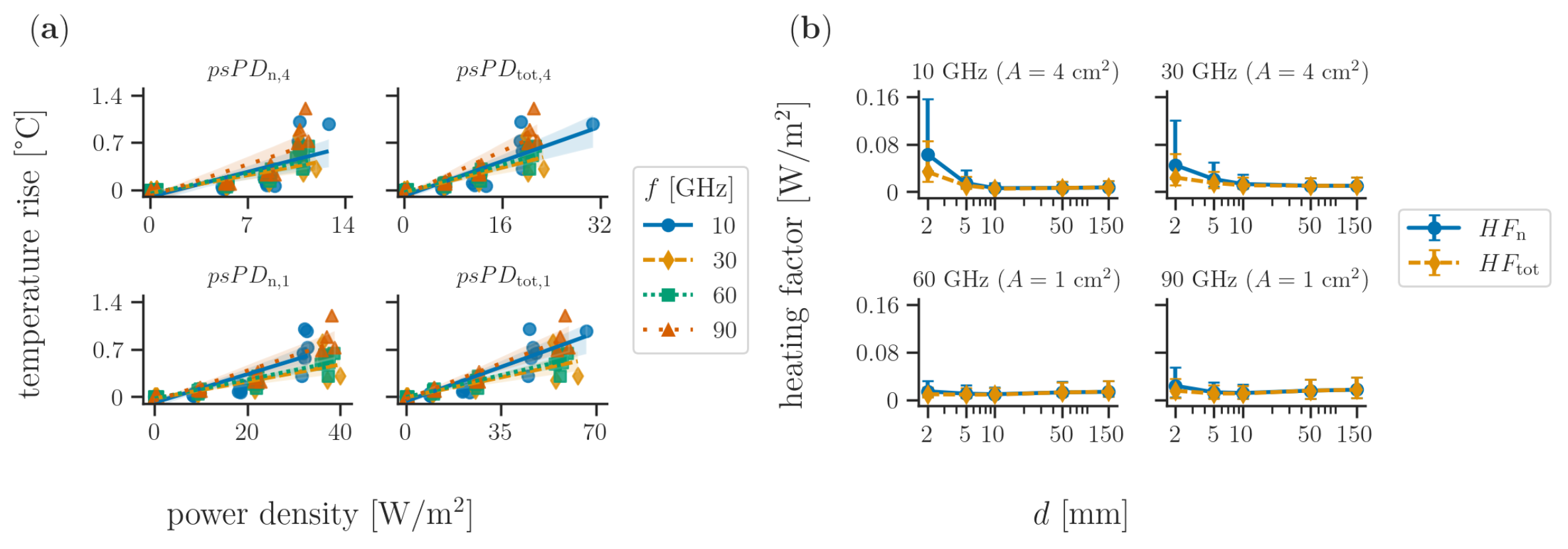}
        \caption{Relationship between maximum temperature rise and different power densities incident on the surface shown as: (a) linearly dependent (b) indirectly, through the heating factor as a function of the antenna-to-tissue separation distance.}
        \label{fig:2}
    \end{figure}

    In~\cref{fig:2}(a), the correlation between maximum temperature rise and the corresponding spatially averaged \gls{ipd} is shown.
    In all cases, the Pearson correlation coefficient is greater than 0.7.
    The linear dependence of maximum temperature rise and the \gls{ipd} on the surface is therefore also displayed indirectly by using the heating factor in~\cref{fig:2}(b).
    At \SIlist{10;30}{\GHz}, a significant increase in all heating factors is observed at \SI{2}{\mm}.
    However, at $d > \SI{10}{\mm}$, heating factors considering either definition of the spatially averaged \gls{ipd} are constant within the margin of error with respect to $d$.

\subsection{Development of the Synthetic Dataset}
    Considering that surrogate modelling based on data-driven methods requires a large amount of data, the dataset with different combinations of features, i.e., $d$, $f$, $sPD_\text{n, 1}$, $sPD_\text{tot, 1}$, $sPD_\text{n, 4}$ and $sPD_\text{tot, 4}$, have been manufactured artificially by using a statistical modelling approach.
    To this end, the Gaussian copula -- a multivariate normal distribution which captures both the marginal probability distribution of each feature in the input collected data and the dependence structure between them, was fitted.
    Additionally, dosimetric constraints were enforced to the copula to avoid non-physical values during sampling from the fitted model.
    In total, \num{10000} synthetic rows composed of multivariate samples were generated to populate the dataset subsequently used during surrogate modelling.
    All integer-valued frequencies in the \SIrange{10}{90}{\GHz} range and separation distances in the \SIrange{2}{150}{\mm} range were sampled from the uniform distribution.
    For each frequency and distance pair, \gls{ipd}s and resultant maximum temperature rise were sampled from the fitted non-parametric kernel density function which allows smooth and unconstrained approximations. The synthetic dataset is free and open-source and available on GitHub in \verb+thermal-dosimetry-surrogate+ public repository at \url{https://github.com/akapet00/thermal-dosimetry-surrogate.git}.

\subsection{Surrogate Modelling}
    Overall, 4 different surrogate models were fitted in a supervised manner for prediction of maximum temperature rise on the surface for different scenarios of exposure to a half-wavelength dipole antenna.
    The synthetic dataset was split in 80-20 ratio where the first \SI{80}{\percent} of the data was used to fit a model and the resulting \SI{20}{\percent} was used to test each model and optimize its hyperparameters. 
    
    XGBoost~\cite{Chen2016} served as the baseline model due to simple implementation for the regression problem in question and compatibility with the input data format.
    Hyperparameters were optimized by employing the grid search method over pairs of critical values using the 5-fold cross validation splitting strategy and early stopping set to 20 rounds to avoid overfitting.
    
    Neural network models have also been used as surrogate models in this study.
    The first one -- \gls{mlp}, is a fully connected feed-forward neural network with 4-unit input layer, 3 hidden layers with 256 units, and a single-unit output layer.
    All but units of the output layer are ReLu activated.
    The network weights were obtained by using the Adam optimizer with the learning rate fixed to 0.001 over \num{100000} iterations.
    The batch size was set to 128.
    Additionally, TabNet, a deep neural network based on a sequential attention mechanism, was trained over \num{100000} iterations with assistance of self-supervised learning with unlabelled data to improve the selection of relevant features at each training iteration~\cite{Arık2019}.
    The Adam optimizer was used with the adaptive learning rate set to 0.001 at the first iteration.
    
    Lastly, the \gls{moe} model, was compiled as a combination of 3 different surrogate models.
    The first one is the square polynomial model which can be expressed as $\boldsymbol{y} = \boldsymbol{X} \beta + \varepsilon$, where $\boldsymbol{y}$ is the model’s output, $\boldsymbol{X}$ is a design matrix composed of the input data, and $\varepsilon$ is a vector of random errors.
    Then, estimated polynomial regression coefficients were obtained simply by using the closed-form analytical expression, $\beta = \boldsymbol{X}^\intercal \boldsymbol{X}^{-1} \boldsymbol{X}^\intercal \boldsymbol{y}$,~\cite{Hastie2009}.
    The remaining two models are the regularized minimal-energy tensor product B- and cubic Hermite splines, mathematical details available in~\cite{Hwang2018}.
    This hybridized modelling approach increased overall accuracy by replacing a single global model by a weighted sum of each surrogate model which operates on its own cluster of data~\cite{Bouhlel2019}.

\section{Results and Discussions}
\label{sec:results}

\subsection{Synthetic Data Quality and Diagnostic Report}
\begin{figure}[t]
        \centering
        \includegraphics[width=\textwidth]{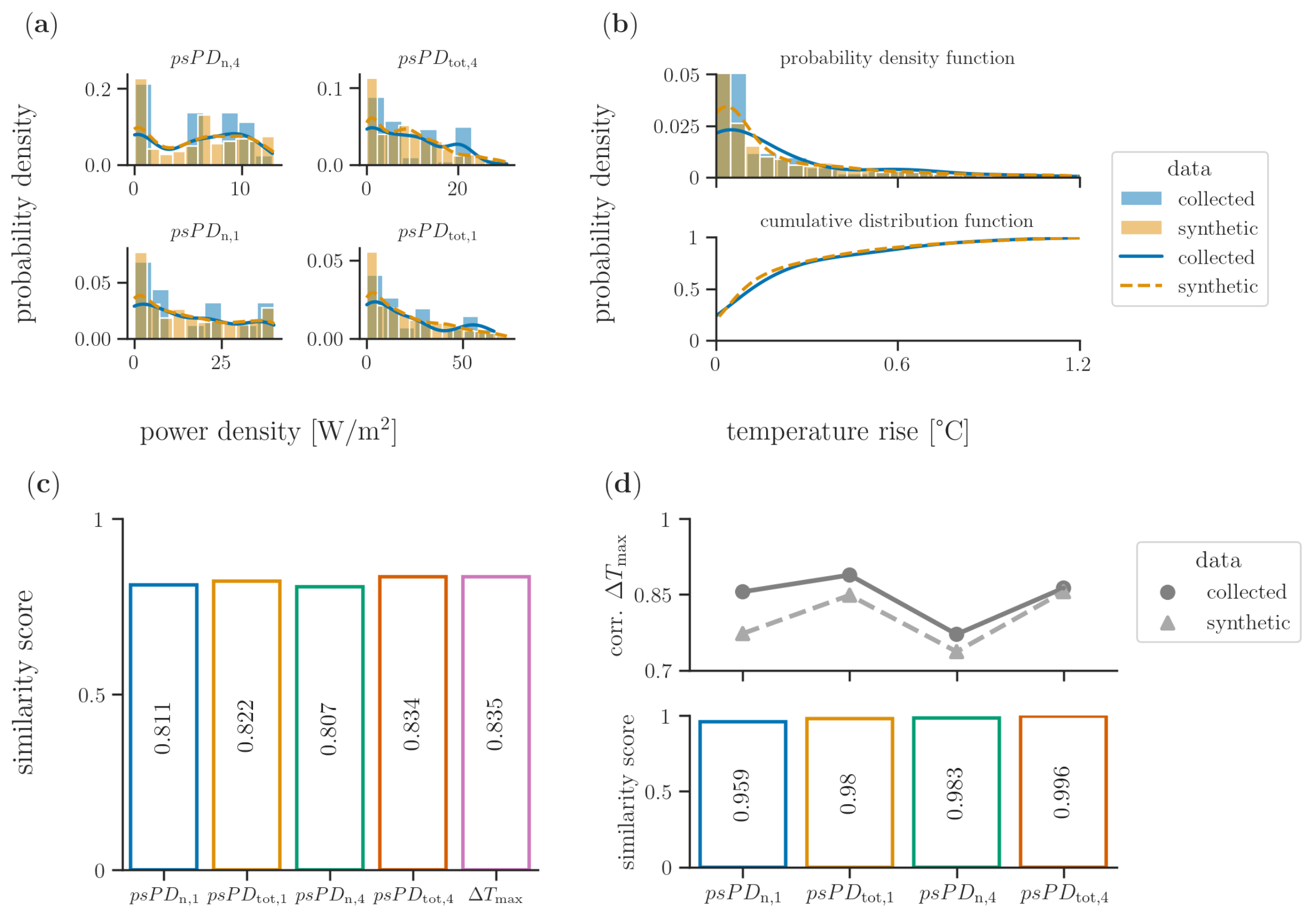}
        \caption{Histograms and estimated marginal distributions of the original (``collected'', full line) and artificially generated data (``synthetic'', dashed line) of: (a) power densities incident on the surface, (b) resulting temperature rise on the surface. Similarity scores of the corresponding: (c) marginal distributions and (d) correlation coefficients in the collected and synthetic data.}
        \label{fig:3}
    \end{figure}

    Marginal distributions of the collected data and their artificially generated synthetic counterparts for spatially averaged \gls{ipd}s and maximum temperature rise on the surface are shown in~\cref{fig:3}{a} and~\cref{fig:3}{b}, respectively.
    
    The similarity of the collected and synthetic data was derived from the \gls{ks} test statistic.
    A 2-sample non-parametric \gls{ks} test quantifies the equality of continuous distribution functions of the collected and synthetic data in terms of the maximum distance.
    The similarity score, shown as a bar chart in~\cref{fig:3}{c}, was then computed as $1 - \text{\gls{ks}}$ test statistic, where a higher score indicates greater equality.
    
    As shown in~\cref{fig:2}, there exists a relatively strong correlation between all peak spatially averaged \gls{ipd}s and maximum temperature rise in the collected data.
    Pearson correlation coefficients were then calculated on the synthetic and compared with those of the collected data, see top panel in~\cref{fig:3}{d}.
    The similarity score of pair trends is reported here as $1 - \text{the normalized relative difference between the corresponding correlation coefficient in the collected and synthetic data}$, and is shown in the bottom panel in~\cref{fig:3}{d}.
    Furthermore, a 2-tailed statistical significance test considering the Fisher $z$-score was carried out to check whether there is a statistically significant difference between the calculated coefficients.
    Results show that in all cases, except for $psPD_\text{n, 1}$, the null hypothesis (correlation coefficient in the collected data is equal to its synthetic counterpart) is retained with the $p$-value greater than predefined threshold significance level of 0.05.
    
    Finally, quantitative diagnostics were measured focusing on the coverage and synthesis.
    In all cases, except for $psPD_\text{tot, 4}$, the synthetic data cover the entire range of possible values found in the collected data.
    About \SI{10}{\percent} of the values of $psPD_\text{tot, 4}$ are outside the boundaries of the collected data and therefore do not provide accurate coverage.
    Furthermore, each row in the synthetic dataset is unique, which indicates the completeness of the synthesis and the absence of all potential issues related to data leakage during surrogate modelling.
    This was confirmed by calculating the proportion of normalized values in the synthetic data that match the collected data within \SI{1}{\percent}, resulting in 0.

\subsection{Evaluation of Surrogate Models}
    All surrogate models have been developed to predict maximum temperature rise considering the antenna-to-tissue separation distance given frequency and spatially averaged \gls{ipd} that satisfied statistical tests outlined in previous section.
    Fitting was done by using the synthetic dataset, where \SI{80}{\percent} of the data was used for the training and the resulting \SI{20}{\percent} for hyperparameter tuning.
    Once fitted, the performance of surrogate models was examined on the collected data and quantified by means of the \gls{mae}.
    
    The results are as follows.
    Overall, the most accurate model is the \gls{moe} with the resulting \gls{mae} of \SI{0.058}{\celsius}.
    This is only a marginal improvement compared to the baseline model with the \gls{mae} of \SI{0.063}{\celsius}.
    However, advantages of the \gls{moe} are the simplicity of the implementation and the training speed in the order of \SI{10}{\s} compared to the order of \SI{e4}{\s} required for training the baseline model, depending on the exhaustiveness of the hyperparameter search.
    For reference, experiments were run on an Ubuntu 22.04.1 LTS machine with the 8-core Intel i5-1135G7 processor and 16 GB of memory.
    
    On the other hand, the \gls{mlp} resulted with the \gls{mae} of \SI{0.091}{\celsius} which is about \SI{50}{\percent} greater compared to the baseline model.
    This is mainly attributed to neural networks generally being biased to overly smooth solutions and not being robust to uninformative input features~\cite{Grinsztajn2022}.
    A better performance is achieved with TabNet owing to its sequential attention structure.
    Regardless, with the \gls{mae} of \SI{0.069}{\celsius}, TabNet is still inferior to the \gls{moe} in terms of performance.

\section{Conclusions}
\label{sec:conclusions}
    In this work, a comprehensive open-source dataset, compiled of artificially generated \gls{ipd}s and resulting temperature rise based on the existing exposure data at \SIrange{10}{90}{\GHz}, was presented.
    This out-of-the-box dataset is primarily intended for computational dosimetry researchers to be able to develop prototypes without wasting resources and risking errors that can occur during \gls{emf} data collection based on computationally intensive simulations.
    To prove its validity, 4 different surrogate models (gradient-boosted trees, quadratic polynomials in combination with tensor-product splines and 2 neural networks) for approximation of temperature rise based on the frequency, antenna-to-tissue separation distance and 3 different \gls{ipd}s were fitted.
    Their performance was tested by using the originally collected data on the basis of which the statistical model from which the synthetic dataset itself was created.
    The \gls{moe} model, a combination of polynomial/tensor-product spline surrogates, achieves overall best results in terms of the \gls{mae} of only \SI{0.058}{\celsius}.
    It outperforms neural network-based surrogate models by up to \SI{50}{\percent} demonstrating the effectiveness of a hierarchical approach where the problem domain is divided into multiple subdomains, each assigned to a different small-scale expert.

\section*{Acknowledgements}
This research was supported by the European Regional Development Fund under the grant KK.01.1.1.01.0009 (DATACROSS).

\printbibliography

\end{document}